\documentclass[12pt,preprint]{aastex}

\newcommand{\etal}{{et al.}\ }
\newcommand{\eg}{{e.g.,}\ }
\newcommand{\ie}{{i.e.,}\ }

\shorttitle{Dusty DAZ White Dwarfs}
\shortauthors{von Hippel \etal}

\begin{document}

\title{The New Class of Dusty DAZ White Dwarfs}

\author{Ted von Hippel\altaffilmark{1,2}, Marc J. Kuchner\altaffilmark{3},
Mukremin Kilic\altaffilmark{4}, Fergal Mullally\altaffilmark{1}, William
T. Reach\altaffilmark{5}}

\altaffiltext{1}{Department of Astronomy, University of Texas at Austin, 1
University Station C1400, Austin, TX 78712-0259; ted@astro.as.utexas.edu}
\altaffiltext{2}{Visiting Scientist, Southwest Research Institute, 1050 
Walnut St., Suite 400 Boulder, CO 80302} 
\altaffiltext{3}{NASA Goddard Space Flight Center, Greenbelt, MD 20771}
\altaffiltext{4}{Columbus Fellow, Department of Astronomy, Ohio State
University, 140 West 18th Avenue, Columbus, OH, 43210}
\altaffiltext{5}{Spitzer Science Center, MS 220-6, California Institute of
Technology, Pasadena, CA 91125}

\begin{abstract}

Our mid-infrared survey of 124 white dwarfs with the {\it Spitzer}
Space Telescope and the IRAC imager has revealed an infrared excess
associated with the white dwarf WD 2115$-$560 naturally explained by
circumstellar dust.  This object is the fourth white dwarf observed to
have circumstellar dust.  All four are DAZ white dwarfs, \ie they have
both photospheric Balmer lines and photospheric metal lines.

We discuss  these four objects as a class, which we abbreviate ``DAZd'',
where the ``d'' stands for ``dust''.  Using an optically-thick,
geometrically-thin disk model analogous to Saturn's rings, we find
that the inner disk edges are at $\ga$ 0.1 to 0.2 $R_\sun$ and that the
outer disk edges are $\sim$0.3 to 0.6 $R_\sun$.  This model naturally
explains the accretion rates and lifetimes of the detected WD disks and
the accretion rates inferred from photospheric metal abundances.

\end{abstract}


\section{Introduction}

The first white dwarf known to have a dusty debris disk or cloud was
Giclas 29-38, discovered by Zuckerman \& Becklin (1987) to have pronounced
excesses in the $K$, $L$, and $M$ bands relative to the white dwarf's
photosphere.  Initially, it was unclear whether the IR excess indicated a
substellar companion (Zuckerman \& Becklin 1987) or particulate debris.
Eventually the near-IR to mid-IR (Tokunaga, Becklin, \& Zuckerman 1990;
Chary, Zuckerman, \& Becklin 1999) spectral energy distribution (SED),
time-resolved photometry in the visible and near-IR that constrained the
dust geometry (Graham \etal 1990; Patterson \etal 1991), and limits on the
presence of companions from pulsation-timing studies (Kleinman \etal 1994)
and speckle imaging (Kuchner \etal 1998), argued in favor of particulate
debris in a disk.  Now new {\it Spitzer} 7--14 $\mu$m spectroscopy
(Reach \etal 2005b) shows a strong 10 $\mu$m emission feature in the G
29-38 excess that can only be caused by small silicate dust particles.
If G 29-38 has any planets, they are currently undetected.

While we do not have strong evidence regarding the distribution of the
dust around G 29-38, we do have a few constraints: the near-IR pulsation
data suggest a non-spherical dust distribution (Graham \etal 1990;
Patterson \etal 1991) and the particles we see in emission radiate at
temperatures of 290 to 890 K (Reach \etal 2005b).  Given the luminosity of
the white dwarf (2 $\times 10^{-3}$ $L_\sun$), these temperatures places
the dust particles at approximately 0.15 to a few solar radii (Zuckerman
\& Becklin 1987; Jura 2003; Reach \etal 2005b); the outer radius depends
on whether we assume the dust is in an optically thin or thick component.

G 29-38 is a relatively normal, H-atmosphere (DA) WD currently passing
through its instability strip, it has $T_{\rm eff} \approx$ 11800 K
(Kepler \& Nelan 1993; Bergeron, Wesemael, \& Beauchamp 1995), log(g) =
7.9 (Koester \& Wilken 2006) to 8.14 (Bergeron \etal 1995), and an implied
mass = 0.56 to 0.69 M$_\sun$.  It has been cooling as a WD for $\sim$0.5
Gyr.  The initial-final mass relation of Weidemann (2000) suggests that
this range of WD masses corresponds to a Zero Age Main Sequence (ZAMS)
mass of 1.2 to 3.1 M$_\sun$.  G 29-38 was therefore once an F or A
star (Drilling \& Landolt 2000, table 15.8).  G 29-38 has photospheric
absorption lines from heavy elements (Ca, Mg, Fe; Koester, Provencal,
\& Shipman 1997), which makes it a spectral type DAZ.  Additionally,
the photospheric Ca II K line varies in strength on timescales as short
as 15 days (von Hippel \& Thompson 2007), indicating episodic accretion.

Becklin \etal (2005) and Kilic \etal (2005) recently discovered a second
white dwarf with circumstellar dust debris.  This object, GD 362, was
found to have a broad-band, $JHKL'M'$ SED that could be fit by a debris
disk (Becklin \etal 2005).  Its near-IR spectroscopy shows a strong
$K$-band excess without any of the near-IR spectral features of a brown
dwarf (Kilic \etal 2005).  GD 362 is an H-rich\footnote{D. Koester,
private communication, reports that GD 362 has a mixed He/H atmosphere
and that the prior mass estimate is substantially too high.  This will
affect the timescale of gravitational settling and thus the implied
accretion rate, but not our analysis.} atmosphere WD with metals observed
in its photosphere; it is in fact the most metal-rich WD known.  It is
slightly cooler than G 29-38 ($T_{\rm eff}$ = 9740 K, Gianninas, Dufour,
\& Bergeron 2004; Kawka \& Vennes 2004), yet is substantially more
massive at 1.24 M$_\sun$ (Gianninas \etal 2004), which placed it at $>$
7 M$_\sun$ on the main sequence (Weidemann 2000).

A third WD with circumstellar dust, GD 56, was just reported by Kilic
\etal (2006b).  This star was discovered via its pronounced excess
at wavelengths longer than 1.6 microns in low resolution near-IR
spectroscopy.  GD 56 is also a DAZ and has $T_{\rm eff}$ = 14400 K and
log(g) = 7.8 (Koester \etal 2005).  These reported atmospheric parameters
correspond to a low mass WD (0.52 M$_\sun$), which is expected only for
a halo star or a remnant from binary star common envelope evolution.
On the other hand, if we assume a slightly higher surface gravity, the
inferred mass becomes normal for a Galactic disk WD, \eg for log(g) =
7.9, mass = 0.566 M$_\sun$ (Bergeron \etal 1995).

In this paper we report the fourth DAZ with circumstellar dust,
WD2115$-$560.  This WD has $T_{\rm eff}$ = 9700 K and log(g) = 8.1
(Koester \etal 2005), corresponding to a mass of 0.66 M$_\sun$.
We discovered this object in our {\it Spitzer} IRAC 4.5 and 8 $\mu$m
survey of 124 WDs (Mullally \etal 2007).  We conducted this survey to
improve upon the ground-based $K$-band survey of Zuckerman and Becklin
(1992) and Farihi, Becklin, \& Zuckerman (2005), and the 6.75 $\mu$m
space-based, ISOCAM survey of Chary \etal (1999) of 11 WDs, none of
which found any new WDs with circumstellar dust.  Our survey also aimed
to study the photospheres of cool WDs (see Kilic \etal 2006a), which
are of great importance to WD cosmochronology, as well as to search for
giant planets and brown dwarfs (see Mullally \etal 2007).

Our discovery of the fourth WD with circumstellar dust leads us to
suggest a new class of white dwarfs, the ``DAZd'' white dwarfs, where
``d'' indicates circumstellar dust.  We choose a lower case ``d'' to
avoid confusion with double white dwarfs, known as double degenerates
and abbreviated ``DD''.  The classification notation for WDs is based
on spectroscopic features, and up to now all have been notated by
capital letters.  Yet these features are all thought to be photospheric,
whereas the dust causing the IR excess is circumstellar; a lower case
``d'' emphasizes the difference between photospheric and circumstellar
spectroscopic properties.

Using the photometry and discovery statistics from our survey, the
detailed properties of G 29-38 reported by Reach \etal (2005b), and some
simple models, we explore the nature and cause of particulate debris in
DAZd WDs as a group.

\section{Observations}

\subsection{{\it Spitzer} Data}

Our {\it Spitzer} survey initially targeted all WDs in the McCook and
Sion (1999, 2003) catalog brighter than $K_S$=15 as measured by 2MASS
(Skrutskie \etal 2006), rejecting known binaries and planetary nebulae,
yielding a total of 134 WDs.  All of these WDs have well-determined
positions and most have well-determined proper motions, facilitating
cross-referencing between catalogs.  We removed one object to avoid a
conflict with the {\it Spitzer} Reserved Observations Catalog and the time
allocation committee removed three other WDs in common with a program
focused on DAZ WDs.  Our observations for six of these WDs failed to
provide quality photometry, mostly because of blending problems, and in
the end, we obtained good data for a sample of 124 WDs with $T_{\rm eff}$
= 5000 to 60,000 K, cooling ages of 1 Myr to 7 Gyr, and representing
most WD spectral types.

Our observations consisted of five dithers of 30 seconds each (150 seconds
total integration time) for every WD in our survey.  We used the products
of the {\it Spitzer} Science Center pipeline, the Basic Calibrated Data
(BCD) Frames and the Post-BCD frames (mosaics), for our analysis.  We used
IRAF\footnote{IRAF is distributed by the National Optical Astronomy
Observatory, which is operated by the Association of Universities for
Research in Astronomy (AURA), Inc., under cooperative agreement with the
National Science Foundation.} PHOT as well as a custom IDL package to
perform aperture photometry on individual BCD frames.  We also attempted
PSF-fitting photometry but found that the poorly defined Point Response
Function for the IRAC instrument with its large pixels (1.213 and 1.220
arcseconds at 4.5 and 8 microns, respectively) meant that we obtained
better results with the IDL and IRAF aperture photometry than with
PSF-fitting photometry.  Both the IDL and IRAF approaches gave identical
results, within the errors.  In order to maximize the signal-to-noise
ratio, we used 5 pixel apertures for bright, isolated objects, and 2
or 3 pixel apertures for faint objects or objects in crowded fields.
We corrected the resultant fluxes by the aperture correction factors
determined by the IRAC team (see the IRAC Data Handbook).  For each
object, we compared the photometry from 2, 3, and 5 pixel apertures;
we found them all to be consistent within the errors.

Following the standard IRAC calibration procedure, we made corrections
for the observed array location of each WD before averaging the fluxes
of the five frames in each IRAC band.  We also performed photometry
on the mosaic images and found the results to be consistent with the
photometry from individual frames.  Based on the calibrations of Reach
\etal (2005a), we expect that our IRAC photometry is calibrated to an
accuracy of 3\%.  We estimated the photometric error from the observed
scatter in the five images (corresponding to 5 dither positions) plus
the 3\% absolute calibration error, added in quadrature.  See Mullally
\etal (2007) for further details of our reduction procedure.

\subsection{Survey Summary}

Our flux-limited survey of 124 single WDs detected circumstellar dust
around two WDs, WD2115$-$560 and G 29-38, which met the criteria for
inclusion in our sample.  These two detections among the general
population of white dwarfs indicate a detection rate of 1.6\% in
our flux limited WD sample.  However, other surveys have shown higher
circumstellar dust detection rates with more limited samples.  Kilic \etal
(2006b) found the third circumstellar dust WD candidate, GD 56, in a
small survey of DAZ WDs; they estimated that $\sim$9\% of all DAZs have
circumstellar dust detectable at $K$.  These two samples differ both in
wavelength coverage and target selection.  The {\it Spitzer} survey was
more sensitive to cooler circumstellar dust.  The IRTF survey of Kilic
\etal (2006b) selected WD targets with high atmospheric metal content,
the DAZs.  The greater sensitivity to WDs with circumstellar dust in the
{\it Spitzer} survey did not yield more such systems; the DAZ survey at
the IRTF had a $\sim$5.6 times higher yield.  This result, along with
the fact that our new circumstellar dust system, WD2115$-$560, turned
out to be a DAZ (a fact we did not know when we assembled the sample),
suggests a deep connection between detectable circumstellar dust around
WDs and the DAZ phenomenon.

Table 1 summarizes the properties of WD2115$-$560, along with the three
other WDs hosting circumstellar dust already presented in the literature.
These WDs span a modest temperature range of 9700 to 14400 K, though their
mass range is wide for WDs, running from $\sim$0.56 to $\sim$1.24 M$_\sun$.
This mass and $T_{\rm eff}$ range mean their cooling ages range from
$\sim$0.2 to $\geq$ 2.5 Gyr.  All four display photospheric metal lines.
We list the temperature range of their circumstellar dust from a simple
model fit, which we discuss below.

Rounding out the known properties of these WDs, Becklin \etal (2005)
and Reach \etal (2005b) find that the fractional luminosity over all
wavelengths in the dust is $\sim$3\% in GD 362 and G 29-38.  The minimum
dust mass around G 29-38, assuming an optically-thin cloud of dust
that radiates efficiently in the thermal infrared, is $10^{18}$ g,
approximately equivalent to the mass of a 10 km asteroid.

\subsection{Blackbody Models for the Spectral Energy Distributions}

Perhaps G 29-38, the first known and best studied DAZd, is a good template
for understanding this emerging class.  Using archival IUE spectra,
optical photometry from the literature, and {\it Spitzer} IRAC, IRS,
and MIPS data, Reach \etal (2005b) fit an optically thin dust cloud
model to the G 29-38 SED.  Reach \etal (2005b) showed that a simple
model consisting of two modified blackbodies, with temperatures of 290
and 890 K, provided a good fit to the infrared excess.

Their {\it Spitzer} spectra revealed a strong 10 $\mu$m emission feature,
well fit by a combination of forsterite, Mg$_2$SiO$_4$, and olivine,
(Mg,Fe)$_2$SiO$_4$.  The assumptions of an optically-thin dust cloud and
blackbody dust place the detectable dust at 1 to 5 $R_\sun$ from the WD.
The Poynting-Robertson (P-R) timescale for particles to spiral onto the
WD from a debris reservoir at 1 $R_\sun$ is only 4 $a$ years, where $a$
is the particle radius in microns.  The silicate emission feature observed
by Reach \etal (2005b) can only be as strong as it is relative to the
continuum (contrast = 125\%), if it is emitted by submicron particles.
The P-R timescale for such tiny particles is years or less.

Radiation pressure can eject small grains in orbit around a star
(\eg Wyatt \& Whipple 1950).  A spherical blackbody grain with radius
$r_{grain}$ and density $\rho_{grain}$ (in g cm$^{-3}$) will be ejected
from bound orbit around a WD by radiation pressure if $r_{grain} <
r_{blowout}$, where
\begin{equation}
r_{blowout} = 1.15 \mu\rm{m} \ \rho_{grain} \ {{L_{WD}}\over{L_{\odot}}} \ 
{{M_{\odot}}\over{M_{WD}}}\ .
\end{equation}
However, for a typical WD luminosity, $L_{WD} = 10^{-3} L_\sun$, and
WD mass, $M_{WD} = 0.6 M_{\odot}$, the blowout size is 0.002 $\mu$m,
much smaller than the grains producing the 10 $\mu$m feature.

\subsection{Flat Disk Models for the Spectral Energy Distributions}

An optically-thin dust cloud would have a lifetime no longer than the
short P-R time scale.  Therefore, it seems likely that the dust cloud is
optically thick, shielding much of the dust from the stellar radiation
that causes P-R drag.  With this idea in mind, we fit the SEDs of each
of the four WD debris systems using standard optically-thick flat disk
models (Friedjung 1985; Jura 2003).  We derived the input spectra for
these simple disk models by fitting blackbody curves to the photospheric
portions of the SEDs, and then used the models to derive disk inclinations
and dust temperatures that correspond to the inner and outer radii of
the disks.

For each WD, we created a grid of models with inner disk temperatures
$T_{\rm in}$ = 600 to 1250 K, outer disk temperatures $T_{\rm out}$ = 200
to 1175 K, $T_{\rm in} > T_{\rm out}$, and a range of inclination angles.
We then found the model that best fit the IR photometry of each WD (J
band and beyond) by minimizing $\chi^2$.  Table 1 lists the temperature
ranges and inclinations of the best fit models.

For GD 56 and GD 362 we included the IRTF spectroscopy of Kilic \etal
(2005, 2006b) in the data to be fit.  For G 29-38 we performed one
fit using our {\it Spitzer} photometry and another fit using the IRTF
(Tokunaga \etal 1999) plus ISO photometry (Chary \etal 1999), though
without the 10.5 $\mu$m photometry, which is likely to be influenced
by the strong silicate emission feature.  The fit to the older IRTF and
ISO photometry was poorer than the fit to the {\it Spitzer} data, and we
report only the fit to the modern data in Table 1.  The difference between
the {\it Spitzer} and older data sets most likely reflects calibration
errors that plagued the ground-based and particularly the ISO photometry
(R. Chary and B. Zuckerman, private communication).  It is also possible
that the G 29-38 debris disk has evolved substantially over the 15 year
span of these observations.

Figures 1--4 present our best fit debris disk models.  All four DAZd WDs
show a pronounced IR excess relative to the expected WD photospheres,
represented by the WD model atmospheres (solid lines, kindly provided
by Detlev Koester)  The debris disk components (dashed lines) fit the
near-IR and mid-IR excess well.

Table 1 also lists the inner and outer radii of the best fit flat disk
models.  They range from 0.15 to 0.58 $R_\sun$ = 13 to 49 white dwarf
radii for the three normal mass WDs, and 0.08 to 0.50 $R_\sun$ = 17 to
106 white dwarf radii for GD 362, assuming the high mass value currently
in the literature, which implies a much smaller radius.  The disk model
has no explicit inner edge directly exposed to radiation from the WD,
and thus the inner radii are significantly underestimated.

\section{A Possible Physical Description of the Disks}

Perhaps the observations of DAZ disks all point to a common simple
picture originally suggested by Jura (2003); the disks are analogs of
planetary rings, physically thin disks of rocks and dust with optical
depth $\approx$ 1.  Our new data make this analogy much more compelling.
As shown above, the physically thin, optically-thick disk geometry
implied by this picture can match the observed SEDs of DAZd WDs.
Besides providing an excuse for that disk geometry, the planetary-ring
analogy has the following virtues.

\noindent 
1. This picture naturally incorporates both small dust grains and larger,
long-lived source particles.

\noindent 
Standard models of planetary rings (\eg Cuzzi \etal 1979) contain a
distribution of particle sizes.  The largest ring particles contain
most of a ring's mass.  These particles occupy a ``monolayer'' a few
particle-widths thick, while the smaller particles---like the ones
observed to produce the 10-micron emission feature in G29-38 (Reach
\etal 2005b)---are spread over a thicker poly-layer.  Planetary rings
may also consist of dusty rings interspersed with rings of larger bodies
(\eg de Pater \etal 2006).  The small particles are continually produced
by the collisions of the larger particles, and protected to some degree
from radiation forces by the optical thickness of the ring.

\noindent 
2.  This picture provides a disk lifetime consistent with the ages of
the DAZd WDs.

\noindent 
An optically-thick disk of rocks has a lifetime set by the viscous
spreading time of the ring, $t_{visc} \approx r^2/\nu$, where $r$ is the
radius of the ring and $\nu$ is the kinematic viscosity.  To estimate the
disk lifetime, let us assume that the viscosity of the WD rings is similar
to that of Saturn's rings.  The viscosity of a monolayer ring is $\nu
=\Omega \tau R_{\nu}$, where $\tau$ is the optical depth and $R_{\nu}$
is a representative particle size, weighted toward the larger particles
(Goldreich \& Tremaine 1982).  The angular speed in the WD ring is similar
to that in planetary rings:  $\Omega \approx 2 \times 10^{-4}$ s$^{-1}$.
So assuming a similar viscosity for the WD debris disks and Saturn's rings
is like assuming a similar optical depth and particle size distribution.
With this assumption, the WD ring lifetimes are $\sim$100 times longer
than the lifetime of Saturn's rings because of their larger radii.
This lifetime is easily $\gtrsim 10^9$ years, comparable to the median
post-main sequence lifetime of the WDs in our survey.  In this picture,
WD rings do not require continual replenishment from a longer-lived
source of small bodies.

\noindent 
3. This picture implies a metal accretion rate consistent with the
photospheric abundances of metals in the DAZd WDs.

\noindent 
In the context of this picture, we can estimate the mass of a typical
DAZd disk as $\pi r_{out}^2 \Sigma$, using the surface density of
Saturn's rings ($\Sigma \approx 40$ g cm$^{-2}$; Tiscareno \etal 2007)
and an outer radius $r_{out} \approx 1.5\ R_\sun$ (see item 5, below).
With these assumptions, we find a representative disk mass of $\sim$2
$\times 10^{-4} M_{\bigoplus}$.  If the accretion time is $10^{9}$ years,
the resulting accretion rate onto the WD is $\sim$7 $\times 10^{-19}$
M$_\sun$ yr$^{-1}$.

Figure 5 shows the observed photospheric abundances derived by Koester \&
Wilken (2006) for a large sample of DAZ WDs.  Figure 6 shows accretion
rates for this sample inferred from these photospheric abundances and
the atmospheric residence times Koester \& Wilken (2006) derived for
these WDs.  These accretion rates from Koester \& Wilken (2006) are
based on an assumption that the accreted material has Solar composition.
Our model assumes that the material is mostly refractory, so to compare
our accretion rate to Figure~6, we need to multiply it by a factor of 50,
giving us $\sim$3 $\times 10^{-17}$ M$_\sun$ yr$^{-1}$.  This accretion
rate sits right in the middle of the range of DAZ inferred accretion
rates, and is only about an order of magnitude less than that of the
known DAZds.  It is easy to imagine a disk with slightly higher surface
density or viscosity than Saturn's rings that would precisely match the
DAZd accretion rates in Figure 6.

\noindent 
4. This picture explains the inner radii of DAZd disks inferred from
their SEDs.

\noindent
The radius where the equilibrium temperature of a dust grain exceeds
the sublimation temperature, $T_{\rm sub}$, for the grains should mark
the inner boundary of the ring.  Although most of the disk sees only
stellar radiation at an oblique angle, the inner radius of the disk can
see direct stellar radiation, so it should be hotter than the flat-disk
model would predict, closer to the temperature of an optically-thin dust
cloud, or a single grain in free space.  With this geometry in mind,
we find that for blackbody grains, the sublimation radius is
\begin{equation}
R_{sub, BB} = \frac{1}{2}\ \left(\frac{T_{sub}}{T_{\sun}}\right)^{-2}
              \left(\frac{L_{WD}}{L_{\sun}}\right)^{1/2} R_{\sun},
\end{equation}
where $L_{WD}$ is the WD luminosity.  For a typical WD luminosity, $L_{WD}
= 10^{-3} L_{\sun}$, and assuming a silicate sublimation temperature,
$T_{\rm sub} = 1500$~K, Equation~2 gives $R_{\rm sub} = 0.24\ R_\sun$.

Another component of the disks may be grains too small to efficiently emit
at the blackbody peak wavelength.  These grains are slightly hotter than
blackbody grains at the same location in the nebula, so they sublimate
at larger radii.  E.g., for the ``medium''-sized grains described by
Backman \& Paresce (1993), the sublimation radius is 
\begin{equation}
R_{sub, M} = \left(\frac{T_{sub}}{4009.21}\right)^{-5/2} 
             \left(\frac{L_{WD}}{L_{\sun}}\right)^{1/2}
             \lambda_0^{-1/2} R_{\sun},
\end{equation}
where $\lambda_0$ is the grain size measured in microns.  For the same
assumed WD luminosity and silicate sublimation temperature, and for a
grain size of 1~$\mu$m, $R_{\rm sub, M} = 0.37\ R_\sun$.

Table 1, column 12 lists dust sublimation radii for each of the
DAZds calculated from Equation~2 assuming $T_{\rm sub} = 2000$~K.
This assumption produces a good match to the inner radii inferred from
our models of the SEDs.  However, we did not incorporate the physics
of direct heating of the inner rim of the disk into our SED modeling,
so we cannot compare sublimation radii calculated using the assumption
of direct heating to the radii inferred from our flat disk models.
We can, however, compare the inner temperatures inferred from our flat
disk models to the sublimation temperatures of typical dust materials and
retain self-consistency.  Table 1, column 9 lists the temperatures of the
inner and outer edges of the disks inferred from the flat disk models.
The temperatures at the inner edge range from 800--1150~K. 

\noindent 
5. This picture explains the outer radii of DAZd disks inferred from
their SEDs.

\noindent 
A dynamically cold ring cannot extend much beyond the Roche radius;
beyond this radius, the ring particles coagulate into moons---or in the
case of the white dwarf rings---asteroids or planets.  The Roche radius
for tidal disruption of an asteroid near a WD is given by Jura (2003):
\begin{equation}
R_{\rm Roche} = C_{\rm tide}\ R_{\rm WD}\
                \left(\frac{\rho_{\rm WD}}{\rho_{\rm asteroid}}\right)^{1/3},
\end{equation}
where $C_{\rm tide}$ is a parameter of order unity that depends on the
asteroid's composition, orbit, and rotation, $R_{\rm WD}$ is the radius
of the WD, $\rho_{\rm WD}$ is the density of the WD, and $\rho_{\rm
asteroid}$ is the density of the asteroid.  For G 29-38, $R_{\rm Roche}
\approx 126\ C_{\rm tide}\ R_{\rm WD} \approx 1.5\ C_{\rm tide}\ R_\sun
\approx 1.5\ R_\sun$.  Since this equation only includes density to the
one-third power and WD masses span a narrow range ($\sim$0.55 to 1.2
M$_\sun$), all WDs have about the same Roche radii.

Jura (2003) pointed out that the outer radius of the disk around G29-38
roughly matched the Roche radius for this WD.   We find that the outer
radii of all the DAZd disks as inferred from physically-thin, optically
thick disk models for the SEDs are consistent with the WD Roche radii.

\noindent 
6. This picture may explain the absence of dust around luminous
DAZ white dwarfs.

\noindent 
This {\it Spitzer} survey and the survey of Kilic \etal (2006b) examined
eleven DAZ white dwarfs hotter than 15,000 K and found circumstellar dust
around none of them (see Figures 5 and 6).  The high surface abundances
and very short atmospheric residence times for calcium (see \S 4) in
these eleven stars made them excellent candidates to host observable
debris disks.  G\"ansicke \etal (2006) discovered a single WD with
effective temperature $\sim$22,000 K that hosts a calcium gas disk,
but shows no sign of circumstellar dust.  These observations suggest
that perhaps hotter, more luminous WDs cannot host dust.

The analogy between WD disks and planetary rings suggests a natural
interpretation of these observations; for hot WDs, the sublimation zone
reaches beyond the Roche radius.  These WDs, like main sequence stars,
lack a circumstellar region where small bodies are stable against both
sublimation and coagulation in dynamically-cold disks.

We can find the critical WD effective temperature above which this
stable region disappears by setting $R_{\rm sub, BB}=R_{\rm Roche}$,
assuming blackbody dust grains:
\begin{equation}
T_{crit, BB} = \sqrt{2\ C_{\rm tide}}\
               \left(\frac{\rho_{\rm WD}}{\rho_{\rm asteroid}}\right)^{1/6}
               T_{sub}.
\end{equation}
Ignoring the slow dependence of this expression on the WD density, we find
the critical WD temperature is $T_{\rm crit, BB} \approx 16\ T_{\rm sub}$.
If instead we use the 1~$\mu$m ``medium'' grains of Backman \& Paresce
(1993), we find that the maximum WD temperature for hosting a dust disk is
\begin{equation}
T_{crit, M} = 0.13 \sqrt{\ C_{\rm tide}}\
              \left(\frac{\rho_{\rm WD}}{\rho_{\rm asteroid}}\right)^{1/6}
              T_{sub}^{5/4}.
\end{equation}
so $T_{\rm crit, M} \approx 1.5\ T_{\rm sub}^{5/4}$.

For grains with $T_{sub}=1500$~K, we find $T_{crit, BB}=24,000$~K,
and $T_{crit, M}=14,000$~K.  We suggested above that the inner edges
of the observed DAZd disks match blackbody grains with $T_{sub}=2000$~K.
This combination implies that there should be no disks around WDs hotter
than $T_{crit, BB}=32000$~K, keeping in mind the caveats described above.
The observations described above suggest that the critical temperature
should be roughly 15,000--22,000 K.  These observations broadly
agree with the hypothesis that the grain sublimation temperature and the
Roche radius together limit the kinds of WDs that can host dust disks,
although clearly details remain to be better understood.

\section{The Connection Between WD Debris Systems and the DAZ Phenomenon}

As we mentioned above, the presence of photospheric metals in DAZs implies
ongoing accretion of these metals.  Our physical model for the DAZd disks
provides a natural explanation for this accretion.  Let us examine in
more detail the connection between DAZd disks and the DAZ phenomenon.

White dwarfs are classified as DAZ or DBZ if their optical spectra
contain metal lines along with much stronger hydrogen or helium lines,
respectively.  White dwarfs showing metal lines may also be classified
simply DZ if they are too cool to show hydrogen or helium lines.
Hybrid types, such as DABZ, also exist.

Since the discovery of metal lines in WD atmospheres (see Weidemann
1958; Greenstein 1960), the origin of these metals has been debated
(\eg Lacombe \etal 1983; Aannestad \etal 1993; Dupuis, Fontaine, \&
Wesemael 1993; Zuckerman \& Reid 1998; Zuckerman \etal 2003; Koester \&
Wilken 2006; Kilic \& Redfield 2007).  The high surface gravity of WDs
should cause the denser atoms of the metals to sink in the atmosphere
on timescales of days to millions of years, depending on the atmospheric
constituents, effective temperature, and WD mass.  These timescales are
all much shorter than the ages of most WDs.

Figure 7 shows the timescale for gravitational settling for DA WDs with
log(g) = 7.75 to 9.0, roughly mass = 0.48 to 1.20 M$_\sun$, taken from
Koester \& Wilken (2006).  For comparison, the four DAZd stars are
plotted on this figure based on their atmospheric $T_{\rm eff}$ and
log(g) values.  Figure 7 shows that for the WDs hotter than 11,000 K,
the settling times are substantially less than a year and even for our
coolest, low surface gravity object, WD2115$-$560, the settling time is
still less than 200 years.

These short settling timescales imply that DAZ stars were very recently
accreting heavy elements.  Interstellar accretion, perhaps the most
discussed scenario for supplying the heavy elements, seems to fail by
orders of magnitude for many of the known DAZ WDs (Aannestad \etal 1993;
although see the counter-argument by Koester \& Wilken 2006).  However,
such accretion would be natural if DAZ WDs typically harbor substantial
circumstellar disks.

So far all debris WD systems with circumstellar disks have detectable
photospheric metal lines.  Is it possible that all metal-rich WDs
(DAZ/DBZ/DZ) accrete their metals via debris disks?

One reason answering this question is difficult is that not all WDs with
photospheric metals can be recognized as DAZ/DBZ/DZs.  This point is
made in Figure 5, where the curve indicating a constant Ca equivalent
width of 15 m\AA\ corresponds to much higher Ca abundances for hotter
WDs than for cooler WDs.  It is much more difficult to recognize a
DAZ at higher stellar temperatures.  Figure 6 shows another view of
the same data, easier to grasp visually, but more model-dependent.
The maximum accretion rates necessary to explain the observed abundances
are essentially independent of $T_{\rm eff}$; the short atmospheric
residence timescales for the hot WDs are balanced by the deeper convective
reservoirs for the cooler WDs.

The DAZd WDs are located near the top of Figures 5 and 6; they are
among the WDs with the highest photospheric heavy element abundances and
the highest inferred accretion rates, and therefore should harbor the
most massive and easiest to detect of the debris disks.  It thus seems
possible that the debris disks found to date display an observational
selection effect, and that the majority of DAZs harbor debris disks.

For all DAZd WDs except the possibly massive GD 362, Koester \& Wilken
(2006) find accretion rates in the narrow range of 4.07--8.13 $\times
10^{-16}$ M$_\sun$ yr$^{-1}$.  Koester \& Wilken (2006) assumed that
the accreted material had solar abundances.  We assume instead that
the accreted material lacks hydrogen and helium, reducing the total
accretion rate by a factor of 50.  This lower limit corresponds to 8 to
16 $\times 10^{-18}$ M$_\sun$ yr$^{-1}$ or 1.6 to 3.2 $\times 10^{16}$
g yr$^{-1}$ of refractory elements fed to the WD, a rate consistent
with viscous accretion in our planetary-ring model.  If this accretion
persists for at least 25\% of the lifetime of the WDs, commensurate with
the fraction of DAs that are DAZs (Zuckerman \etal 2003), it corresponds
to an accreted mass of $\ga 7 \times 10^{-4}$ of an Earth mass accreted
over 1 Gyr.

\subsection{Origin of the Debris}

We have discussed the basic properties of the debris systems found to
date and we have connected the debris observable in the infrared to
the photospheric metal lines observable in the optical.  We have argued
that dust debris is common around WDs and that this dust must be steadily
accreting onto the WD.  We have not yet examined possible sources for the
debris around WDs.  Here we consider possible origins:  WD mergers, debris
left during the asymptotic giant branch (AGB) or proto-planetary nebulae
(pPNe) phases, and accretion of dust derived from planetary system bodies.

\subsubsection{White Dwarf Mergers}

Merging WDs are expected to leave behind disks (Livio, Pringle, \& Wood
2005; Hansen, Kulkarni, \& Wiktorowicz 2006) to shed angular momentum.
According to the calculations of Livio \etal (2005), such disks should be
massive ($\sim$0.007 M$_\sun$), extend to 1 AU or more, and predominantly
composed of carbon and oxygen, assuming the merger of two WDs each with
mass $\le$ 0.7 M$_\sun$.  The minimum mass for most of these merged WDs
ought to be $\ge$ 1.1 M$_\sun$, since it takes about 10 Gyr to produce a
0.55 M$_\sun$ WD.

Except for GD 362, with a mass possibly equal to 1.24 M$_\sun$, these
mass constraints rule out this explanation for the majority of DAZd WDs.
The other WDs with circumstellar dust have masses ranging from 0.52 to
0.69 M$_\sun$.  Additionally, the disk mass and radial extent appear
inconsistent with all DAZd stars discovered to date.  While WD mergers
must be taking place, and they may possibly be the source of the disk
around GD 362, they are not the source of the majority of the observed
DAZd systems.

\subsubsection{Late Stages of Stellar Evolution}

The outflows of late-stage asymptotic giant branch (AGB) stars and
proto-planetary nebulae (pPNe) are often highly non-spherical.  This
process is not understood, and a range of models have been proposed for
the origin of the asymmetries, including stellar companions and magnetic
fields (see Balick \& Adams 2002, and references therein).  There is
circumstantial evidence for the formation of disks and tori during the
AGB to post-AGB star transition, albeit at much greater distances from
the host star, and most pPNe are producing or have recently produced dust
particles (Balick \& Adams 2002).  In particular, Waters \etal (1998)
observed olivines and pyroxenes around a pre-white dwarf.  While most
of the outflows are ejected well above the stars' escape velocities, at
present it is not known whether some small fraction of these outflows
remain in orbit after they are ejected.  Remaining stellar material
is a possible explanation for the dust disk discovered by Su \etal
(2007) around (at 35--150 AU) the Helix planetary nebula.  While Su
\etal attribute the observed dust to disrupted Kuiper Belt objects or
disrupted comets, it could also be the result of stellar material ejected
with less than the escape speed.

This scenario predicts dusty environments for the youngest WDs with
less and less dust as WDs age.  The observations, however, run counter
to this.  All four of the known DAZd systems have old host WDs with
ages of $\geq$ 0.2 to $\geq$ 2.5 Gyr and no hot, dusty WDs have been
discovered.  The metal disk WD found by G\"ansicke \etal (2006) pushes
this to slightly younger objects, $\sim$0.1 Gyr, but still, if the late
stages of stellar evolution commonly leave debris disks at a range of
radii, these debris disk should be detectable in our survey and in IUE
observations of hot WDs.  Yet, to date, no such systems have been seen
(Hansen \etal 2006; Kilic \etal 2006b).

\subsubsection{Planetary System Remnants}

The dust around DAZds may represent the remains of planetary systems.
Roughly 5-10\% of A-type stars have substantial debris disks (\eg Backman
\& Paresce 1993) and precise Doppler surveys find giant planets around
roughly 5-10\% of solar-type stars (\eg Fischer \& Valenti 2003).
Perhaps 5-10\% or more of WD progenitors had planetary systems, and
some small bodies in these systems survived the post-main sequence
evolution of their host stars (Debes \& Siguardsson 2002, Jura 2003).
Our observations and models appear to be consistent with this picture
for the origin of all the DAZd disks.

\section{Conclusion}

We presented the properties of WD2115$-$560, a new WD with infrared
excess we discovered in our {\it Spitzer} IRAC photometry, and the fourth
DAZ seen to have an infrared excess well-described by circumstellar
dust debris.  We examined the dusty DAZ WDs as a class, which we refer
to with the letters ``DAZd''.

Using a simple, flat, optically-thick disk model motivated by the geometry
of Saturn's rings, we find that the dust in these four DAZd WDs reside
in disks ranging in inner temperature from $\sim$800 to 1150 K and outer
temperature from $\sim$200 to 725 K, not taking into account the direct
heating of the inner edges of the disks.  The outer disk temperatures are
likely to be lower limits set by the sensitivity of the observations and
the amount of dust radiating at that temperature.  These dust temperatures
imply that the inner disk edges are at $\ga$ 0.1 to 0.2 $R_\sun$ and
the detectable extent of the outer disks are $\sim$0.3 to 0.6 $R_\sun$.
The close circumstellar locations of these debris disks are consistent
with simple sublimation calculations and with the Roche radius for tidal
disruption of an asteroid near a WD (Jura 2003).

We find that, taken together, the DAZd WDs point to a disk model analogous
to planetary rings.  This model naturally explains the disk accretion
rates and lifetimes and the temperature distribution of the dusty WDs.
We favor the interpretation that most of these disks represent planetary
systems remnants.

\acknowledgments

This work is based in part on observations made with the Spitzer Space
Telescope, which is operated by the Jet Propulsion Laboratory, California
Institute of Technology under NASA contract 1407.  Support for this
work was provided by NASA through award project NBR 1269551 issued by
JPL/Caltech to the University of Texas.

\clearpage

\thispagestyle{empty}
\begin{deluxetable}{lcrccccccccc}
\rotate
\tablewidth{0pt}
\tablecaption{White Dwarf and Disk Parameters}
\tablehead{
\colhead{star} & \colhead{SpT} & \colhead{$T_{\rm eff}$} & \colhead{mass} & 
\colhead{ZAMS} & \colhead{WD age} & \colhead{$L_{\rm dust}$} &
\colhead{dist} & \colhead{$T_{\rm dust}$} & \colhead{$R_{\rm dust}$} & \colhead{$i$} &
\colhead{$R_{\rm sub,BB}$}  \\
\phantom{12} (1) & (2) & (3) \phantom{1} & (4) & (5) & (6) & (7) & (8) & (9) & (10) & (11) & (12)}
\startdata
GD 56        & DAZd & 14,400 & 0.52--0.57 & 0.9--1.4 & 0.2--0.4 & 4.2\% & 67 & 1125--525 & 0.21--0.58 & 45 & 0.36 \\
G 29-38      & DAZd & 11,600 & 0.56--0.69 & 1.2--3.1 & 0.5      & 3.5\% & 14 & 1150--725 & 0.15--0.28 & 45 & 0.18  \\
GD 362       & DAZd & 9740   & 1.24       & $>$7.0   & $>$2.5   & 6.1\% & 24 &  800--200 & 0.08--0.50 & 60 & 0.07 \\
WD2115$-$560 & DAZd & 9700   & 0.66       & 2.8      & 0.9      & 0.9\% & 22 &  900--550 & 0.17--0.32 & 80 & 0.16  \\
\enddata
\tablecomments{WD and ZAMS masses are in solar units.  WD cooling ages
are in Gyr.  Distances are in pc.  The dust radii and sublimation
radii (see Equation~2) are in units of the solar radius; the sublimation
radii assume $T_{\rm sub}$ = 2000 K.  Mass and cooling ages derived
here use the quoted atmospheric parameters and Bergeron \etal (1995)
tabulated models.  The percentage of luminosity in the disk is the
value from our model fits.  For G 29-38 and for GD 362, Reach \etal
(2005b) and Becklin \etal (2005) estimate 3\% for the disk components
of these two WDs, respectively.  Distances are from Farihi \etal (2005);
Van Altena, Lee, \& Hoffleit (1995); Gianninas \etal 2004; and Gliese \&
Jahreiss (1991), respectively.  Please see text for references to the
atmospheric parameters.}
\end{deluxetable}

\clearpage

\figcaption{The debris disk model and observations for GD 56.  Closely
spaced black dots represent the 0.8 to 2.5 micron spectrum obtained at
the IRTF (Kilic \etal 2006b).  The larger black dots with error bars
typically smaller than the symbols size represent the $UBV$ photometry
from McCook \& Sion (2003) and the $JHK$ photometry from 2MASS (Skrutskie
\etal 2006).  The dashed line represents our fitted debris disk model,
with $T_{\rm in}$ = 1125 K and $T_{\rm out}$ = 525 K.  For comparison
purposes, the solid line represents a WD model atmosphere with $T_{\rm
eff}$ = 14,000 and log(g) = 8.0, kindly provided by Detlev Koester.}

\figcaption{Similar to Figure 1, but for G 29-38.  More extensive IR
photometry are available for this object.  The ``x'' symbols represent the
IRTF photometry of Tokunaga \etal (1990; data obtained during 1987--1989).
The squares represent the ISO photometry obtained by Chary \etal (1999;
observed December 11, 1997).  Our Spitzer data are represented by the
open circles.  The error bars in the IRAC bands are smaller than the
symbol size.  We do not plot the near-IR spectra from IRTF to avoid
crowding in this figure.  Our debris disk model for G 29-38 has $T_{\rm
in}$ = 1150 K and $T_{\rm out}$ = 725 K.}

\figcaption{Similar to Figure 1, but for GD 362.  The $L'$ and $N'$
photometry were obtained at Gemini (Becklin \etal 2005).  We do not use
the $N'$ photometry in deriving the debris disk parameters since it was
likely influenced by the silicate emission band.  Our debris disk model
for GD 362 has $T_{\rm in}$ = 800 K and $T_{\rm out}$ = 200 K.}

\figcaption{Similar to Figure 1, but for WD2115$-$560.  Optical photometry
from McCook \& Sion (2003), near-IR photometry from 2MASS (Skrutskie \etal
2006), and our IRAC photometry for WD2115$-$560 fit a debris disk model
with $T_{\rm in}$ = 900 K and $T_{\rm out}$ = 550 K.  The lower overall
disk flux is best fit by a nearly edge-on disk; $i \approx 80$ degrees.}

\figcaption{$T_{\rm eff}$ vs.\ log(Ca/H) for DAZs from the sample of
Koester \& Wilken (2006).  The solid diagonal curve is the approximate
lower limit calcium abundance for discovery from Koester \& Wilken, at an
equivalent width of 15 m\AA.  The dashed vertical line is the approximate
dividing stellar $T_{\rm eff}$ at which blackbody dust inside the tidal
destruction radius of the WD will be completely sublimated for $T_{\rm
sub}$ = 1200 K.  Other sublimation temperatures and dust properties move
this dividing line a few thousand degrees (see text).  The open symbols
with dots in their centers are the known DAZd WDs.  Besides the four DAZd
WDs discussed here, we also plot an additional DAZd just found by Kilic \&
Redfield (2007) via IRTF spectroscopy.  The filled circles are DAZs with
no detected dust excesses in either the IRTF study of Kilic \etal (2006b)
or in our Spitzer observations.  The remainder (plotted with asterisks)
of the DAZs have not yet been searched for debris disks.}

\figcaption{$T_{\rm eff}$ vs.\ the log of the accretion rate in solar
masses per year, assuming the accretion of solar abundance material, from
Koester \& Wilken (2006).  The symbols are the same as in Figure 5, though
GD 362, with log(dM/dt) $\approx -12.09$, is not plotted for clarity.}

\figcaption{Timescale for gravitational settling of calcium (Koester \&
Wilken 2006).  The location of the four DAZd WDs are indicated by the
open squares.  All have short settling times.}

\clearpage
\epsscale{.90}
\begin{figure}[!t]
\plotone{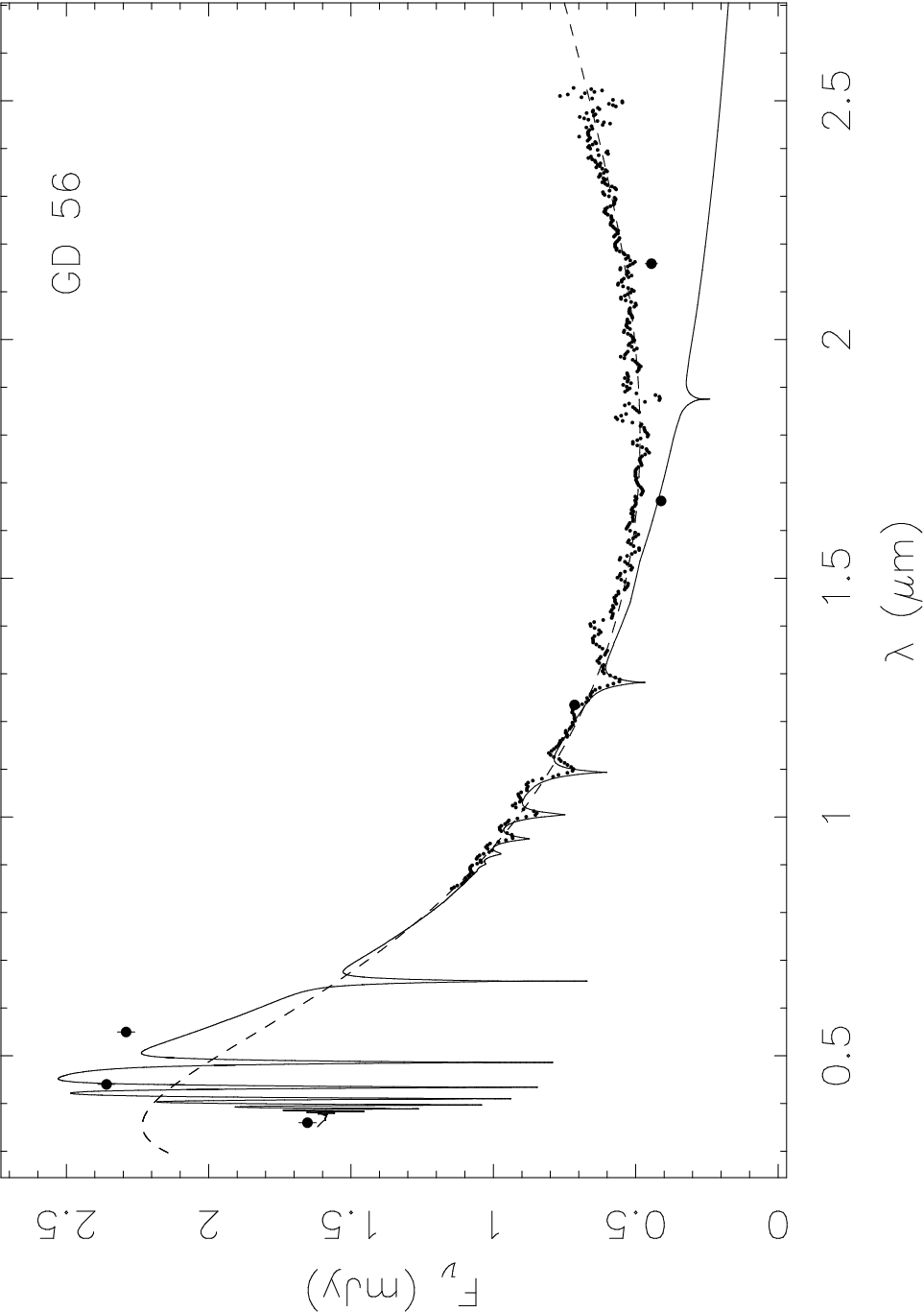}
\centerline{f1.eps}
\end{figure}

\begin{figure}[!t]
\plotone{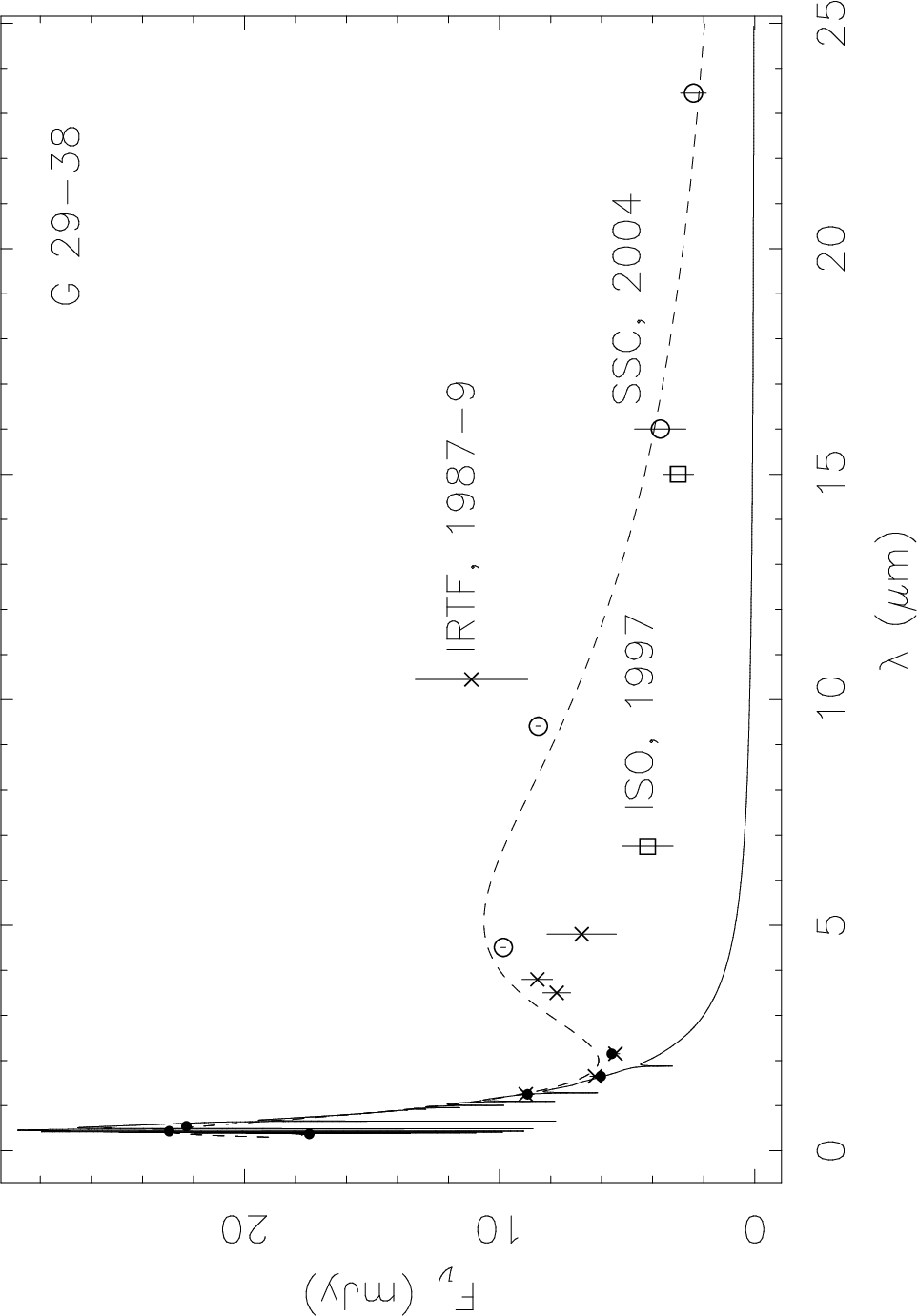}
\centerline{f2.eps}
\end{figure}

\begin{figure}[!t]
\plotone{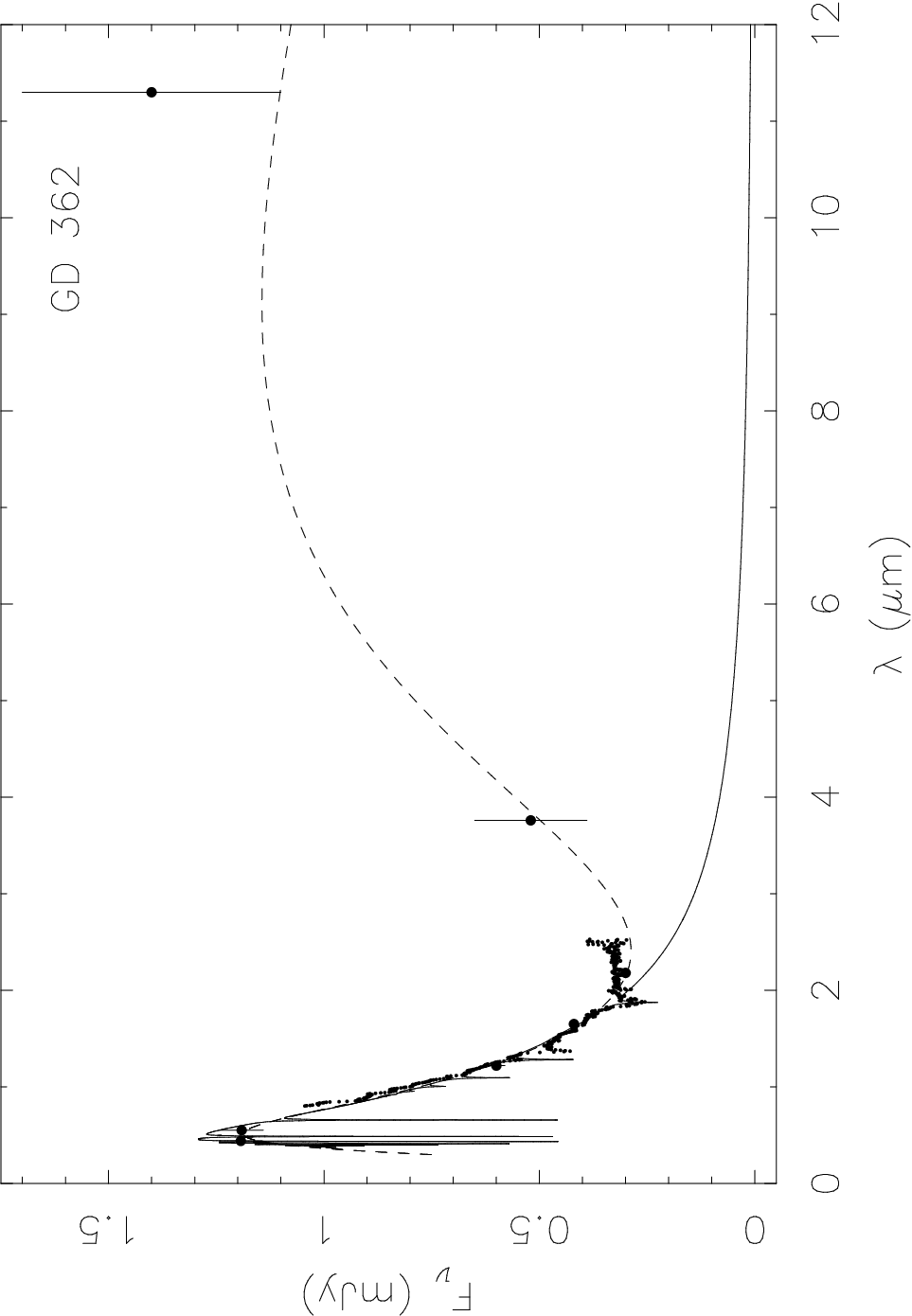}
\centerline{f3.eps}
\end{figure}

\begin{figure}[!t]
\plotone{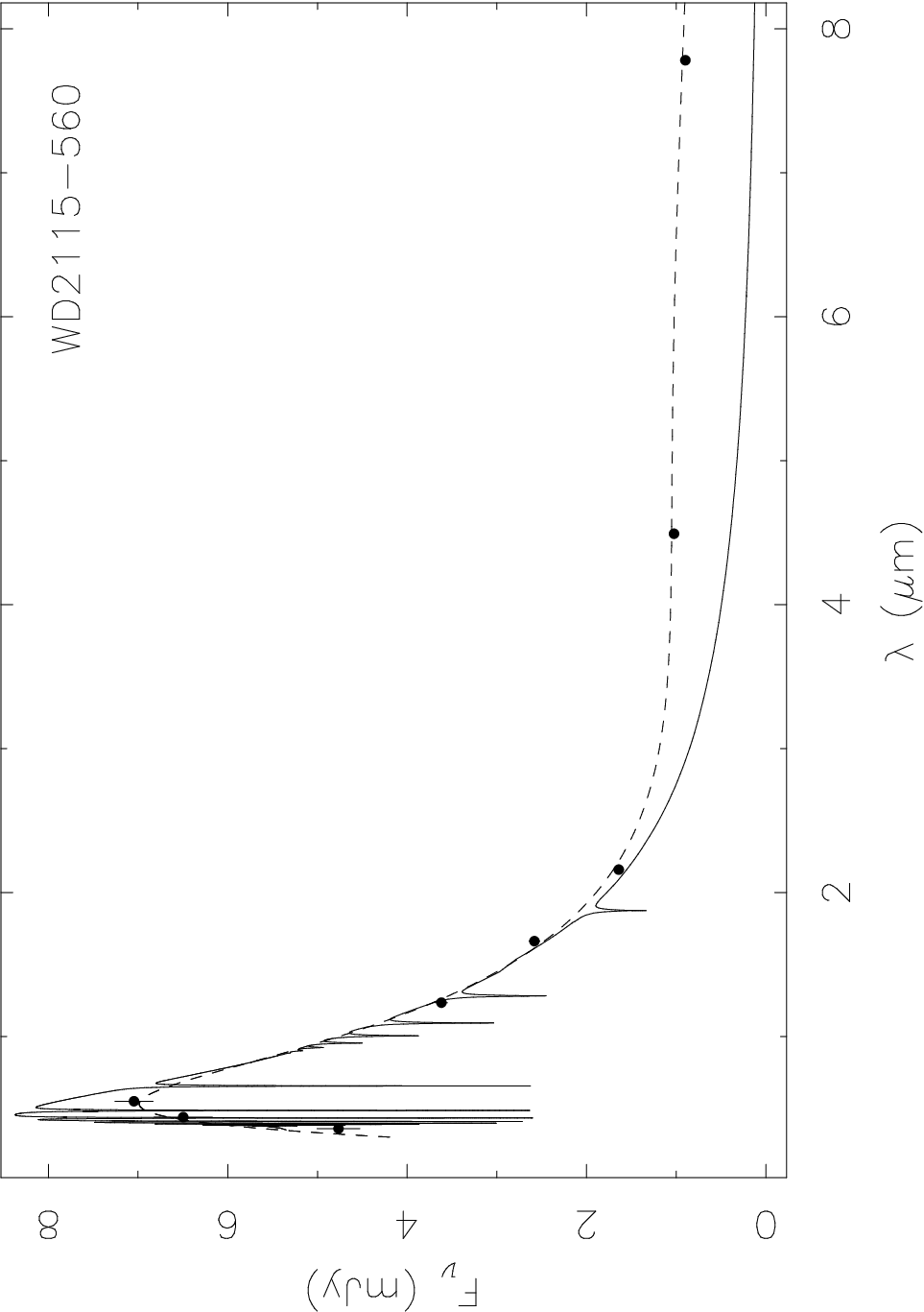}
\centerline{f4.eps}
\end{figure}

\begin{figure}[!t]
\plotone{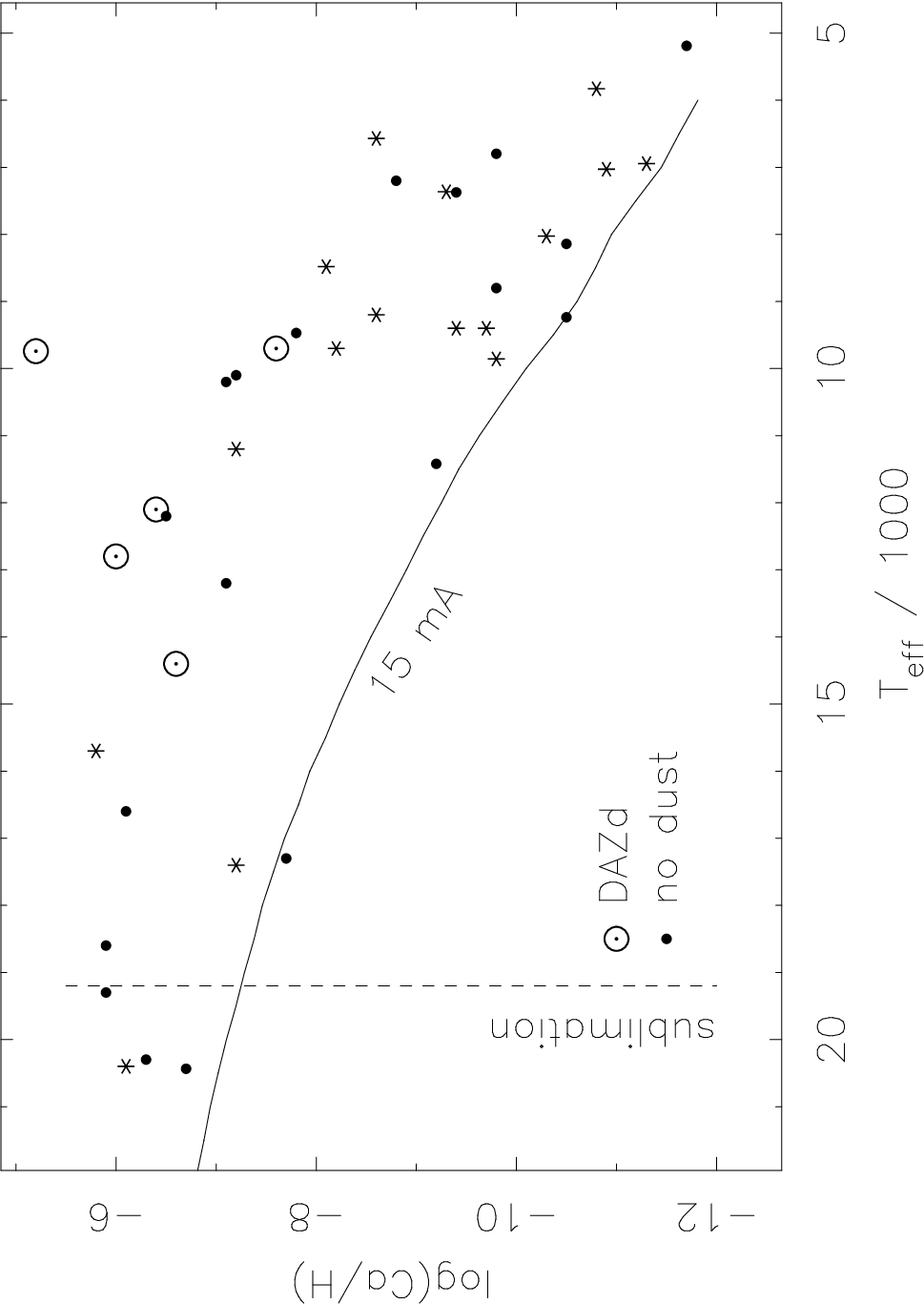}
\centerline{f5.eps}
\end{figure}

\begin{figure}[!t]
\plotone{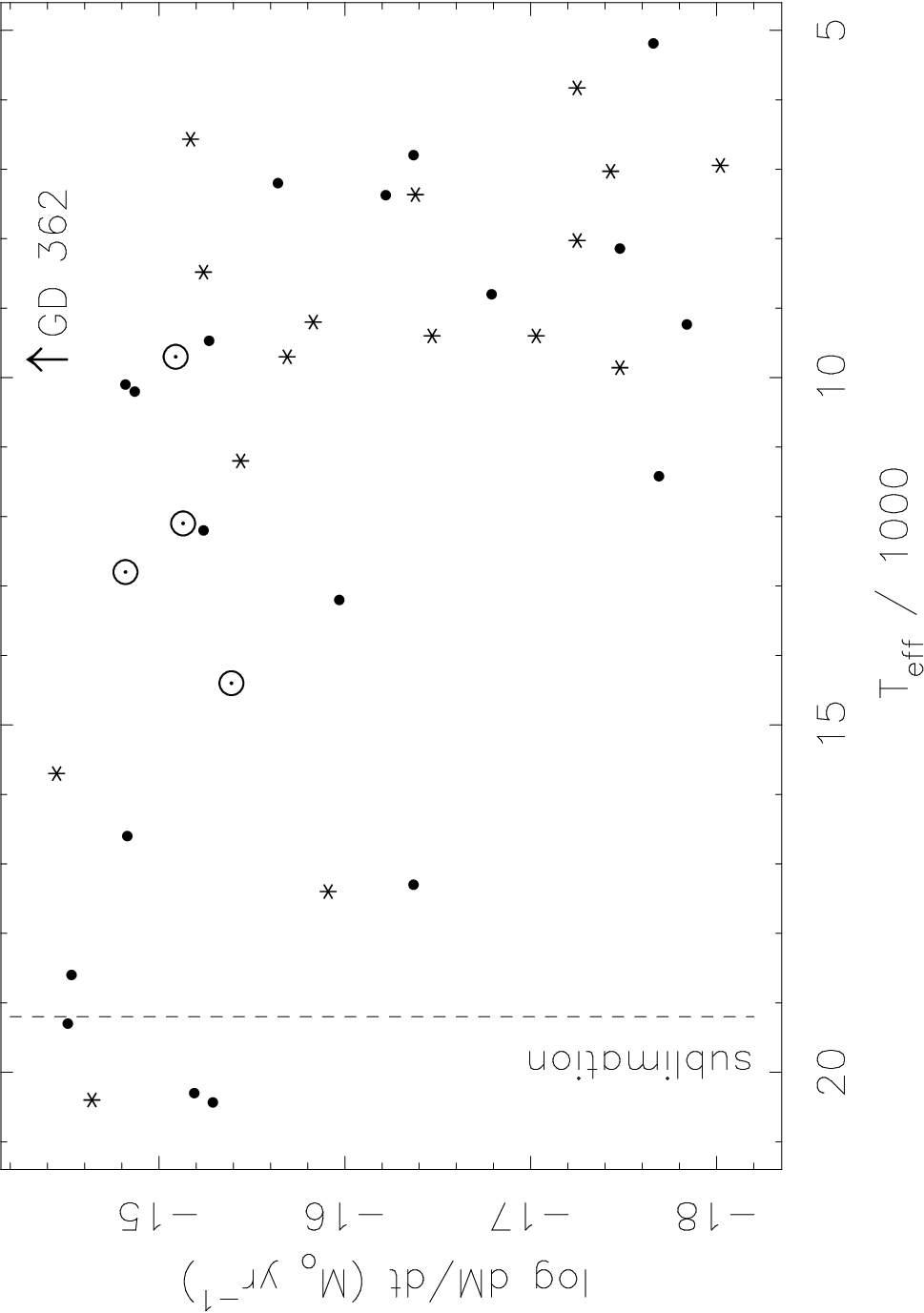}
\centerline{f6.eps}
\end{figure}

\begin{figure}[!t]
\plotone{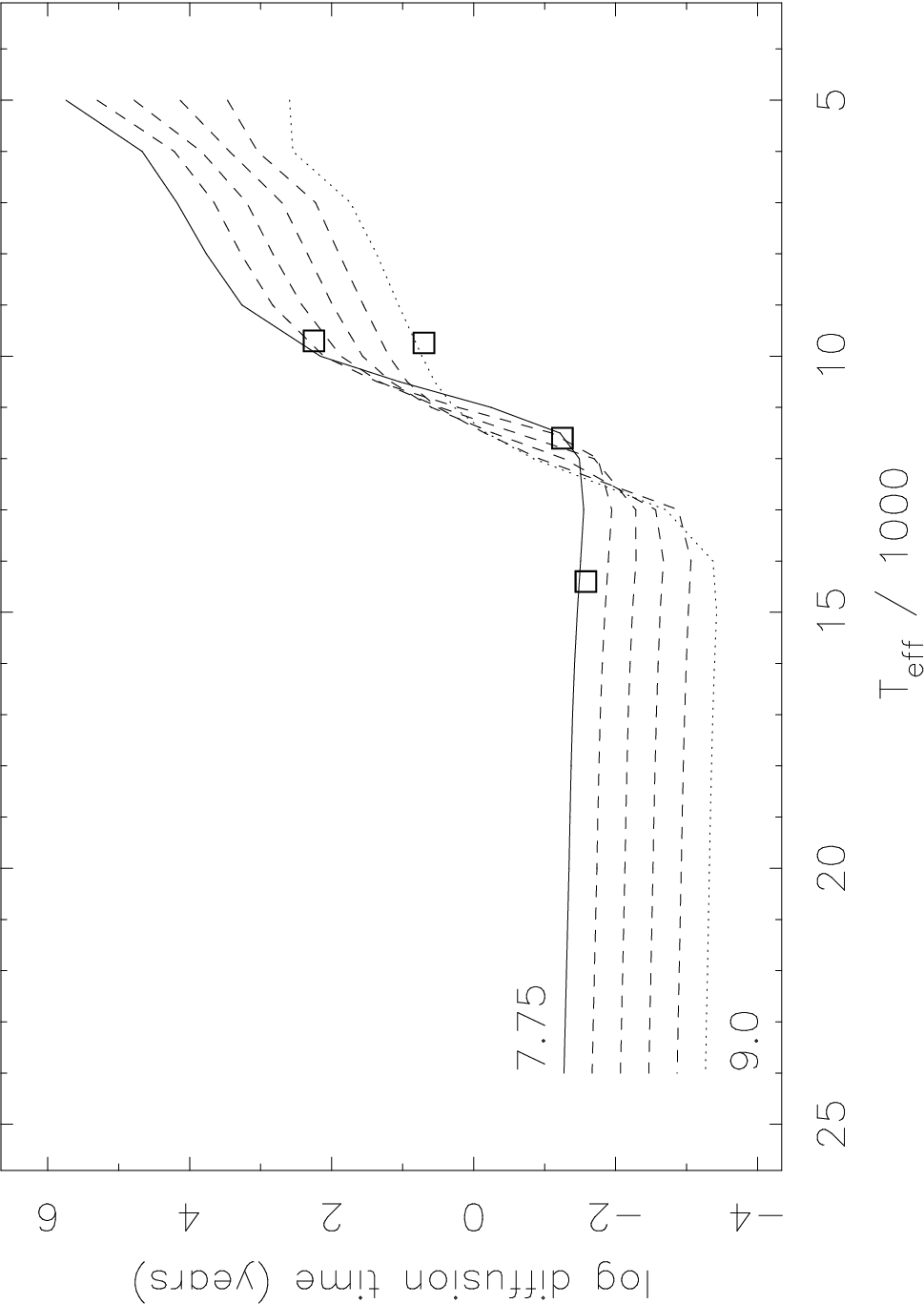}
\centerline{f7.eps}
\end{figure}


\begin{references}

\reference{}
Aannestad, P. A., Kenyon, S. J., Hammond, G. L., \& Sion, E. M. 1993,
\aj, 105, 1033

\reference{}
Backman, D. E. \& Paresce, F. 1993, in Protostars and Planets III,
Main-sequence stars with circumstellar solid material - The Vega
phenomenon, p 1253

\reference{}
Balick \& Adams 2002, \araa, 40, 439

\reference{}
Becklin, E. E., Farihi, J., Jura, M., Song, I., Weinberger, A. J., \&
Zuckerman, B.  2005, \apj, 632, L119

\reference{}
Becklin, E. \& Zuckerman, B. 1988, \nat, 336, 656

\reference{}
Bergeron, P., Wesemael, F., \& Beauchamp, A. 1995, \pasp, 107, 1047

\reference{}
Chary, R., Zuckerman, B., \& Becklin, E. E. 1999, in The Universe as Seen
by ISO, ed. P. Cox \& M. F. Kessler (ESA-SP 427; Garching: ESA), 289

\reference{}
Cuzzi, J. N, Burns, J. A., Durisen, R. H. \& Hamill, P. M. 1979, \nat,
281, 202

\reference{}
Debes, J. \& Siguardsson, S. 2002, \apj, 572, 556

\reference{}
de Pater, I., Hammel, H. B., Gibbard, S. G., \& Showalter, M. R. 2006,
Science, 312, 92

\reference{}
Drilling, J. S., \& Landolt, A. U. 2000, in Allen's Astrophysical
Quantities, Fourth Edition, ed. A. N. Cox, (Springer-Verlag: New York),
p. 389

\reference{}
ibid, p. 395

\reference{}
Dupuis, J., Fontaine, G., \& Wesemael, F. 1993, \apjs, 87, 345

\reference{}
Farihi, J., Becklin, E. E., \& Zuckerman, B. 2005, \apjs, 161, 394

\reference{}
Friedjung, M. 1985, \aap, 146, 366 

\reference{}
G\"ansicke, B.T., Marsh, T.R., Southworth, J., \& Rebassa-Mansergas,
A. 2006, Science, 304, 1908

\reference{}
Gianninas, A., Dufour, P., \& Bergeron, P. 2004, \apj, 617, L57

\reference{}
Gliese, W., \& Jahreiss, H., 1991, Catalog of Nearby Stars,
(Astron.\ Rech.\ Inst., Heidelberg)

\reference{}
Goldreich, P. \& Tremaine, S. 1982, \araa, 20, 249

\reference{}
Graham, J. R., Matthews, K., Neugebauer, G., \& Soifer, B. T. 1990,
ApJ, 357, 216

\reference{}
Greenstein, J. L. 1960, in Stars and Stellar Systems, vol. 6, Stellar
Atmospheres, ed. J. L. Greenstein, (Chicago: University of Chicago Press),
p. 692

\reference{}
Hansen, B. M. S., Kulkarni, S., \& Wiktorowicz, S. 2006, \aj, 131, 1106

\reference{}
Jura, M. 2003, \apj, 584, L91

\reference{}
Kepler, S. O., \& Nelan, E. P. 1993, \aj, 105, 608

\reference{}
Kilic, M., \& Redfield, S. 2007, \apj, in press

\reference{}
Kilic, M., von Hippel, T., Leggett, S. K., \& Winget, D. E. 2005, \apj,
632, L115

\reference{}
Kilic, M., von Hippel, T., Leggett, S. K., \& Winget, D. E. 2006b, \apj,
646, 474

\reference{}
Kilic, M., von Hippel, T., Mullally, F., Reach, W. T., Kuchner, M. J.,
Winget, D. E., Burrows, A., \& Saumon, D. 2006a, \apj, in press

\reference{}
Kleinman, S. J. \etal 1994, \apj, 436, 875

\reference{}
Koester, D., Provencal, J., \& Shipman, H. L. 1997, \aap, 320, L57

\reference{}
Koester, D., Rollenhagen, K., Napiwotzki, R., Voss, B., Christlieb, N.,
Homeier, D., \& Reimers, D. 2005, \aap, 432, 1025

\reference{}
Koester, D., \& Wilken, D. 2006, \aap, 453, 1051

\reference{}
Kuchner, M.~J., Koresko, C.~D. \& Brown, M.~E. 1998, \apj, 508, L81

\reference{}
Lacombe, P., Wesemael, F., Fontaine, G., \& Liebert, J. 1983, \apj, 272, 660

\reference{}
Livio, M., Pringle, J. E., \& Wood, K. 2005, \apj, 632, L37

\reference{}
McCook, G. P. \& Sion, E. M. 1999, \apjs, 121, 1

\reference{}
McCook, G. P. \& Sion, E. M. 2003, VizieR On-line Data Catalog: III/235

\reference{}
Mullally, F., Kilic, M., Reach, W. T., Kuchner, M. J., von Hippel, T.,
Burrows, A., \& Winget, D. E. 2007, \apj, in press

\reference{}
Patterson, J., Zuckerman, B., Becklin, E. E., Tholen, D. J., \& Hawarden,
T.  1991, \apj, 374, 330

\reference{}
Skrutskie, M. F., \etal  2006, AJ, 131, 1163

\reference{}
Reach, W. T., \etal 2005a, \pasp, 117, 978

\reference{}
Reach, W. T., Kuchner, M. J., von Hippel, T., Burrows, A., Mullally, F.,
Kilic, M., \& Winget, D. E. 2005b, \apj, 635, L161

\reference{}
Tiscareno , M. S., Burns, J. A., Nicholson, P. D., Hedman, M. \& Porco,
C. C. 2007, Icarus, in press, (astro-ph/0610242)

\reference{}
Tokunaga, A. T., Becklin, E. E., \& Zuckerman, B. 1990, \apj, 358, L21

\reference{}
Van Altena W. F., Lee J. T., \& Hoffleit E. D. 1995, The General
Catalogue of Trigonometric Stellar Parallaxes, Fourth Edition (Yale
University Observatory)

\reference{}
von Hippel, T., \& Thompson, S. E. 2007, ApJ, in press

\reference{}
Waters, L. B. F. M., \etal 1998, \nat, 391, 868

\reference{}
Weidemann, V. 1958, \pasp, 70, 466

\reference{}
Weidemann, V. 2000, \aap, 363, 647

\reference{}
Wyatt, S.~P., \& Whipple, F.~L. 1950, \apj, 111, 134

\reference{}
Zuckerman, B. \& Becklin, E. E. 1987, \nat, 330, 138

\reference{}
Zuckerman, B. \& Becklin, E. E. 1992, \apj, 386, 260

\reference{}
Zuckerman, B., Koester, D., Reid, I. N., \& H\"unsch, M. 2003, \apj, 596,
477

\reference{}
Zuckerman, B., \& Reid, I. N. 1998, \apj, 505, L143

\end{references}
\end{document}